\documentclass[%
reprint,amsmath,amssymb,aps,
]{revtex4-2}

\usepackage{graphicx}
\usepackage{dcolumn}
\usepackage{bm}
\usepackage{float}
\usepackage{amsmath}
\usepackage[toc,page]{appendix}
\usepackage{blindtext}
\usepackage{multirow,booktabs}
\usepackage{dcolumn}
\usepackage{bm}
\usepackage{units}
\usepackage[version=4]{mhchem}
\usepackage{booktabs,chemformula}
\usepackage{color}
\usepackage[bottom]{footmisc}
\usepackage{xcolor}

\begin{document}

\title{Resistive Switching Conducting Filament Electroformation\\ with an Electrothermal Phase Field Method}

\author{John F. Sevic}
\email{Corresponding author: sevicj@erau.edu}
\affiliation{Department of Electrical, Computer, and Software Engineering, Embry-Riddle Aeronautical University, Prescott, AZ, USA}
\author{Nobuhiko P. Kobayashi}
\affiliation{Department of Electrical and Computer Engineering, University of California, Santa Cruz, CA, USA}

\date{\today}

\begin{abstract}
A phase field method self-consistently coupled to continuum heat transport and charge conservation is used to simulate conducting filament dynamical evolution and nanostructure of electroformed resistive switching thin films. Our method does not require a pre-defined idealized conducting filament, as previous methods do, instead treating its dynamical evolution as a stochastic diffuse interface problem subject to a variational principle. Our simulation results agree well with available experimental observations, correctly reproducing electroformed conducting filament nanostructure exhibited by a variety of resistive switching thin films.
\end{abstract}


\maketitle

Appropriately prepared nanoscale resistive switching thin films exhibit persistent conductivity modulation, central to their operation as next-generation non-volatile memory and neuromorphic computing technologies \cite{Strukov2008}\cite{Waser2007}. Persistent conductivity modulation is produced by various physical mechanisms, such as insulator-metal phase transition (IMT), a consequence of intrinsic metastable atomic-scale states of these thin films and controlled introduction of reversible states endowed by specific preparation processes. 

Various qualitative models have been proposed to study transport phenomena and nanostructure of these thin films, in particular the widely adopted conducting filament (CF) formalism. The CF formalism suggests an initial irreversible forming process producing a nanoscale filamentary thread of locally high electrical conductivity embedded in these thin films. Following CF formation, persistent electrical conductivity modulation obtains by reversible rupture and recovery, usually under the influence of an electric potential and associated Joule heating. While a precise quantitative description of CF dynamical evolution is often not established, the CF formalism nevertheless provides a useful starting point for further quantitative treatment because of experimental evidence suggesting the presence of such conducting filaments.

Building on the CF formalism, many computational studies have adopted a continuum drift-diffusion (DD) formulation of various self-consistently coupled transport phenomena, often referred to as multiphysics \cite{Xu2008}\cite{Pan2010}\cite{Ielmini2011}\cite{Larentis2012}\cite{Nardi2012}\cite{Kim2015}\cite{Sevic2017}\cite{Sevic2018}. In this formulation, a nanoscale CF geometry is \textit{a priori} defined, embedded in an idealized host thin film. The CF formalism then stipulates allied transport phenomena involving coupling between various physical processes and conservation laws. The DD formulation thus requires mobility and diffusion expressions for each specific mass transport mechanism. Additionally, thermal transport and charge conservation are self-consistently coupled.

Supporting the CF formalism is a large body of experimental data providing evidence of the existence of nanoscale conducting filaments in thin films \cite{Yang2009}\cite{Strachan2010}\cite{Miao2011}\cite{Sun2014}\cite{Ahmed2018}\cite{Sun2020}\cite{Zhang2021}. In contrast to the assumption of an \textit{a priori} defined CF with a specific  geometry, these experimentally observed conducting filaments suggest the presence of a stochastic component, due to inherent anisotropic inhomogeneity that naturally appears in thin films. We believe this stochastic character of CF dynamical evolution is central to correctly reproducing experimental nanostructure by computation. Expanding on existing DD formulations, we previously demonstrated an isothermal phase field method requiring no \textit{a priori} assumptions on CF geometry or its idealized thin film host. While making an isothermal assumption, our basic phase field method nevertheless produced conducting filaments exhibiting nanostructure consistent with experimental observation \cite{Sevic2019}.

In this paper, we propose a self-consistent electrothermal phase field method for the computational study of resistive switching phenomena exhibited by a variety of as-fabricated thin films. With this method, the requirement of \textit{a priori} defining a CF and idealized host dielectric is abandoned, and dynamical evolution is instead treated as a stochastic diffuse interface problem subject to a variational principle. A significant feature of the phase field method is that it avoids the mathematically onerous difficulty of expressing dynamic boundary conditions of an unknown diffuse interface, the CF of our method, whose location is part of the solution, while retaining the benefits of the DD formulation.

Our method produces spontaneous nucleation and electroformation of multiple conducting filaments, as seen in a range of resistive switches, offering an alternative computational formulation correctly reproducing experimentally observed nanostructure. Phase change, fundamental to the study of resistive switching based on IMT, is naturally treated by our method, from its metastable atomic origins, and applies to both electronic and ionic charge transport. The present study focuses exclusively on CF electroformation based on ionic transport of a thin film exhibiting unipolar resistive switching from IMT.

\begin{figure}[h]
\centering
\includegraphics[scale=0.22, angle=0]{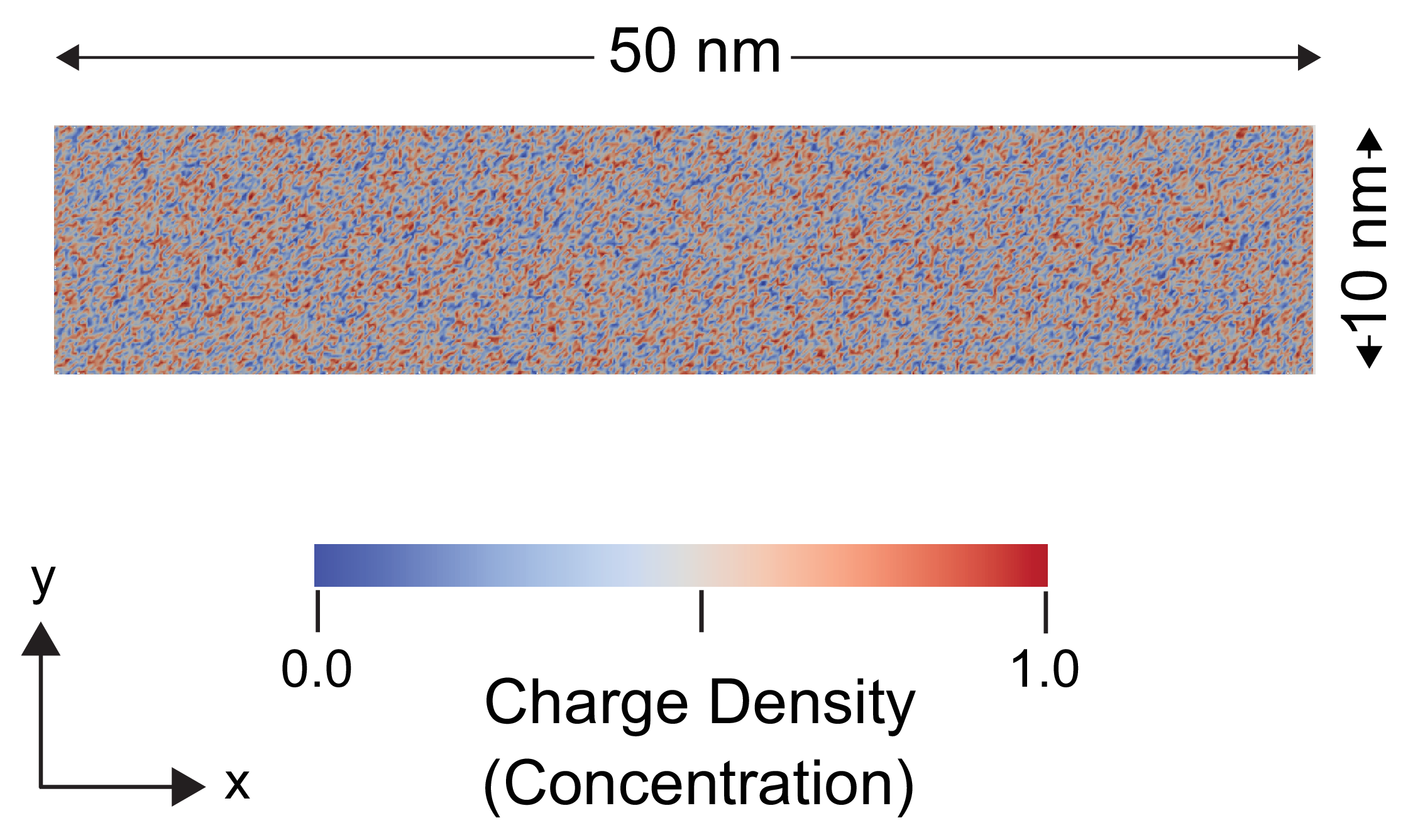}
\caption{Cross-section of our as-fabricated pristine resistive switching thin film model showing dimensions of 50 nm x 10 nm with an initial charge density $c(\vec{r},0)$. The initial charge density is chosen to represent a thin film uniformly at approximately 300 K, corresponding to the free-energy density function of Figure \ref{FEDF}.  To initiate electroformation, an external electric potential of 1 V is applied to the top edge, and the bottom edge is held at 0 V. The top and bottom edges are held at a temperature of 300 K, and periodic boundary conditions along the $x-axis$ are assumed for charge density, electric potential, and temperature. Note that initial charge density is concentrated around the two spinodal minima of the free-energy density function illustrated by Figure \ref{FEDF}.}
\label{DeviceStructure}
\end{figure}

To develop our self-consistent electrothermal phase field method, consider Figure \ref{DeviceStructure} approximating an as-fabricated pristine resistive switching thin film at room temperature with an initial equilibrium charge density $c(\vec{r},0)$. To model resistive switching based on IMT, we assume spinodal separation between an insulating low temperature phase, the pristine equilibrium thin film, and a metallic high temperature phase, the CF of our method \footnote{By pristine we mean pre-electroformed}. From a variational argument, an interface forms, possibly several, under Joule heating because it is energetically favorable, creating isolated clusters of metallic-phase thin film self-consistently evolving, some eventually merging to produce conducting filaments.

Treating IMT from its atomic origins, we adopt a double-well free-energy density based on charge density, $c(\vec{r},t)$, and temperature, $T(\vec{r},t)$, to describe electrothermal CF dynamical evolution producing experimentally consistent nanostructure. This form of free-energy density can approximate a range of resistive switching thin films based on IMT. For the present work we employ an even-order sixth-degree polynomial function of charge density and absolute temperature of the form given by Equation \ref{DoubleWellPotential}

\begin{equation}
\label{DoubleWellPotential}
f_{b}(c,T) = a_{2}(c,T) + a_{4}(c,T) + a_{6}(c,T)
\end{equation}

where $a_{2}(c,T)$, $a_{4}(c,T)$, and $a_{6}(c,T)$ are second-, fourth-, and sixth-degree polynomial functions of normalized charge density and absolute temperature, $c(\vec{r},t)$ and $T(\vec{r},t)$, respectively, and $f_{b}(c,T)$ is free-energy density in keV/mol. Here $\vec{r}$ represents a location in the $(x,y)$-plane of the structure of Figure \ref{DeviceStructure} and $t$ is time \footnote{Spatial and time dependence of the three state variables, $c(\vec{r},t)$, $T(\vec{r},t)$, and $V(\vec{r},t)$, is always implicit, and occasionally may be suppressed subsequently due to space limitations.}.

\begin{figure}[h]
\centering
\includegraphics[scale=0.22, angle=0]{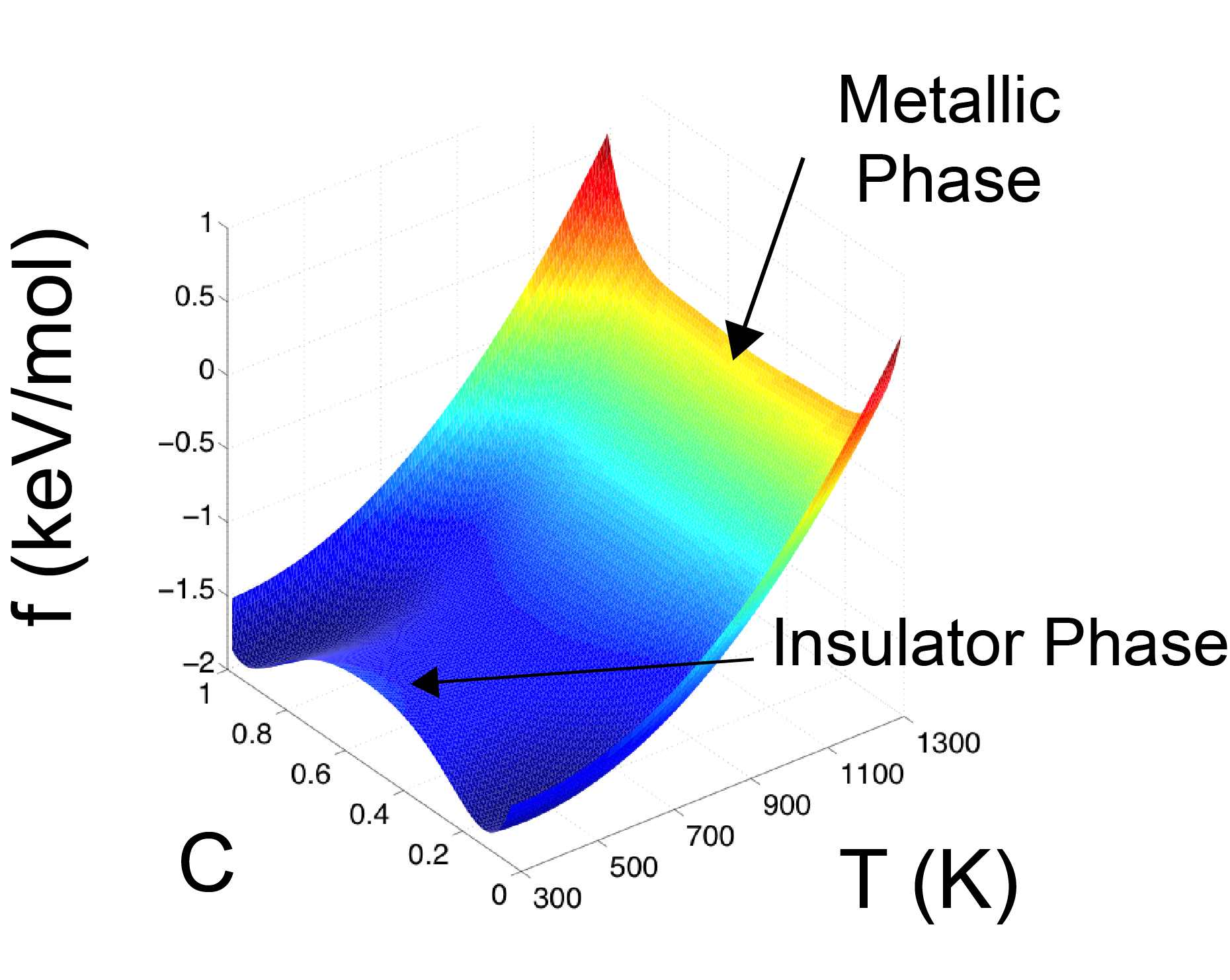}
\caption{Free-energy density as a function of normalized charge density and absolute temperature, $c(\vec{r},t)$ and $T(\vec{r},t)$, respectively. Our phase field method assumes 
spinodal phase separation between an insulating low temperature phase, the pristine thin film, and a metallic high temperature phase, the CF of our method.}
\label{FEDF}
\end{figure}

The constants defining $a_{2}$, $a_{4}$ and $a_{6}$ produce the free-energy density function illustrated by Figure \ref{FEDF}, representing IMT of our simulation. Here the magnitude of the free-energy density and absolute temperature range approximately correspond to thin films from our earlier work \cite{Sevic2017}\cite{Sevic2018}\cite{DiazLeon2017}\cite{DiazLeon2016}.

The free-energy of the thin film and the CF interface energy interact with an externally applied electric potential by an electrostatic energy term

\begin{equation}
\label{ElectricPotentialFreeEnergyDensity}
g_{elec}(c,V) = V(\vec{r},t) \times c(\vec{r},t) \times \frac{q}{\Omega}
\end{equation}

where $V(\vec{r},t)$ is electric potential between the top and bottom edges of the structure, $q$ is electronic charge, and $\Omega$ is a energy density factor that produces Joule self-heating consistent with our IMT model illustrated by Figure \ref{FEDF} \footnote{For the current formulation, $\Omega$ scales the volume of a mesh cell, of unit thcikness.}.

Our phase field CF equation of motion obtains by formulating a free-energy functional composed of Equations \ref{DoubleWellPotential} and \ref{ElectricPotentialFreeEnergyDensity} with the CF interface energy to yield

\begin{equation}
\label{FreeEnergyFunctionalElectric}
F = \int_R \bigg[ f_{b}(c,T) + \frac{\kappa}{2} \nabla^2 c(\vec{r},t) + g_{elec}(c,V) \bigg] \,d\vec{r}
\end{equation}

where $\kappa$ is an interface gradient energy term and the integration is over $R$, the entire thin film of Figure \ref{DeviceStructure}. For the present work, the interface gradient energy term is assumed uniformly constant over $R$. A substantial feature of a variational argument is stationary equilibrium is switching mechanism agnostic. For example, memristive thin films exploiting mechanical strain would be treated by adding to the free-energy functional Equation \ref{FreeEnergyFunctionalElectric} a strain energy density term \cite{Querre2018}\cite{Tranchant2018}.

The equation of motion of the \textit{a priori} unknown diffuse interface, the CF of our method, is found by extracting a Euler-Lagrange equation from free-energy functional Equation \ref{FreeEnergyFunctionalElectric}, yielding

\begin{equation}
\label{PhaseFieldPDE}
\frac{\partial c(\vec{r},t)}{\partial t} =  M\nabla^2 \bigg[ \frac{\partial f_{b}(c,T)}{\partial c} - \kappa\nabla^2 c(\vec{r},t) - \frac{q}{\Omega} V(\vec{r},t) \bigg]
\end{equation}

where $M$ is the CF interface mobility and assumed uniformly constant over $R$. This is a Cahn-Hilliard phase field equation in normalized charge density $c(\vec{r},t)$ for our phase field method \cite{Sevic2019}\cite{Provatas2010}\cite{Cahn1958}. To complete our electrothermal phase field method, this equation of motion governing CF dynamical evolution must be solved self-consistently with the DD heat equation and the charge conservation equation.

Equation \ref{PhaseFieldPDE} is self-consistently coupled to the DD heat equation with a Joule heating source term

\begin{equation}
\label{HeatPDE}
\rho c_{p}\frac{\partial T(\vec{r},t)}{\partial t} - \nabla \cdot k_{th}\nabla T(\vec{r}) = \sigma(\vec{r},c,T)\times \left| \nabla V(\vec{r},t) \right|^{2}
\end{equation}

where $\rho$ is mass density, $c_{p}$ is specific heat capacity, $k_{th}$ is thermal conductivity, and $\sigma(\vec{r},c,T)$ is electrical conductivity at a point in the $(x,y)$-plane of the thin film of Figure \ref{DeviceStructure}. Mass density and specific heat are assumed constant, for the moment, although our electrothermal phase field method naturally treats phase change and anisotropy for these material properties. Thermal conductivity is similarly assumed constant, although its specification is entirely arbitrary with our method.

Charge conservation on normalized charge density, $c(\vec{r},t)$, is imposed as

\begin{equation}
\label{ElectronPDE}
\nabla \cdot\bigg[ \sigma(\vec{r},c,T)\times \nabla V(\vec{r},t)\bigg] - c(\vec{r},t) = 0
\end{equation}

where electrical conductivity, $\sigma(\vec{r},c,T)$, assumes the usual continuum form

\begin{equation}
\label{elecCond}
\sigma(\vec{r},c,T)  = c(\vec{r},t) \times \mu_{F}(\vec{r},T) \times q
\end{equation}

where $c(\vec{r},t)$ is normalized charge density and $q$ is electronic charge. For our electrical conductivity model, we assume the metallic-phase thin film enclosed by the CF interface responds dynamically to electric field $\vec{E}(\vec{r},t) = \nabla V(\vec{r},t)$ with Poole-Frenkel mobility 

\begin{equation}
\label{driftActEnergy}
\mu_{F}(\vec{r},t) = \frac{\mu_{o}}{T(\vec{r},t)} \times exp \left[ \frac{-E_{ac}^{\mu}}{k_{b} \times T(\vec{r},t)} \right]
\end{equation}

where $\mu_{o}$ is a charge mobility pre-factor, $k_{b}$ is the Boltzmann constant and $E_{ac}^{\mu}$ is a field-assisted activation energy.

Our simulation thermal and electrical material properties and phase field model parameters are defined by Table I. We assume isotropic continuum heat transport with mass density, specific heat capacity, and thermal resistance, $\rho$, $c_{p}$, and $k_{th}$, respectively. These thin films exhibit charge transport approximated as field-assisted Poole-Frenkel phenomenon, defined by a charge mobility pre-factor and activation energy, $\mu_{o}$ and $E_{ac}^{\mu}$, respectively. Each of these material properties are extracted from our previous simulated and experimental data on memristive thin films based on IMT \cite{Sevic2017}\cite{Sevic2018}\cite{Sevic2019}\cite{DiazLeon2017}\cite{DiazLeon2016}. The phase field interface mobility, $M$, is approximated to reflect an interface that forms essentially transparent to the CF interface. The interface energy density $\kappa$, is established by making the metal-insulator diffuse interface width consistent with experimental data and our isothermal phase field method \citep{Sevic2019}\cite{KKS}.

\begin{table}[h]
\label{MaterialTable}
\caption{Thermal and electrical material properties and phase field model parameters used electroformation simulation. These parameters appear in our previous experimental and computational work on memrisrive thin films based on IMT.}
\begin{ruledtabular}
\begin{tabular}{ccccc}
\addlinespace[0.5em]
Parameter & Value & Units \\
\addlinespace[0.5em]
\hline
\addlinespace[0.6em]
$\rho$ & 1000 & $\dfrac{kg}{m^{3}}$  \\
\addlinespace[0.8em]
$c_{p}$ & 5.0 & $\dfrac{J}{K \times kg}$  \\
\addlinespace[0.8em]
$k_{th}$ & 0.10 & $\dfrac{W}{m \times kg}$  \\
\addlinespace[0.8em]
$\mu_{o}$ & 100 & $\dfrac{nm}{V \times ns}$  \\
\addlinespace[0.8em]
$E_{ac}^{\mu}$ & 250 & $meV$  \\
\addlinespace[0.8em]
$M$ & 1000 & $\dfrac{nm}{J \times ns}$ \\
\addlinespace[0.8em]
$\kappa$ & 1.0 & $\dfrac{eV}{nm^{2}}$ \\
\addlinespace[0.6em]
\end{tabular}
\end{ruledtabular}
\end{table}

To simulate CF electroformation using our electrothermal phase field method, we solve self-consistently the CF equation of motion, Equation \ref{PhaseFieldPDE}, with the heat equation and charge conservation equation, Equations \ref{HeatPDE} and \ref{ElectronPDE}, using the Multiphysics Object-Oriented Simulation Environment (MOOSE) multiphysics solver \cite{gaston2009moose}. These equations are discretized by an adaptive meshing algorithm and self-consistently solved by a finite element transient Newton method to yield versus time the dynamical evolution of the unknown diffuse interface, $c(\vec{r},t)$, the CF of our method, and state variables $V(\vec{r},t)$ and $T(\vec{r},t)$ \cite{Tonks2010}\cite{schwen2017}\cite{petsc}\cite{libMeshPaper}.

The initial condition for normalized charge density, illustrated by Figure \ref{DeviceStructure}, is chosen to represent an as-fabricated pristine resistive switching thin film, in thermodynamic equilibrium between the two spinodal normalized charge density minima. The initial absolute temperature is uniformly distributed between 300 K and 301 K over the entire thin film. We have found that inhomogeneity in these two initial conditions has a profound first-order effect on CF affinity of	 formation and resultant nanostructure, due to the stochastic nature of CF nucleation and dynamical evolution. The stochastic nature is due to not knowing precisely how locally favorable thermodynamic conditions catalyze CF nucleation and evolution. This is consistent with experimental data and is the subject of future research with our phase field method.

To initiate electroformation, an electric potential on the top and bottom edges of our resistive thin film shown in Figure \ref{DeviceStructure} is established at 1 V and 0 V, respectively.  An absolute temperature boundary condition of 300 K is similarly established for the top and bottom edges. A Dirichlet boundary condition is reasonable given the transient duration of electroformation, and is consistent with our previous experimental work IMT-based evaluation structures embedded in thermal vias \cite{DiazLeon2017}. Periodic boundary conditions for normalized charge density, absolute temperature, and electric potential are imposed on the left and right edges of our resistive thin film. A transient simulation is run to reach the electroformed steady-state. The steady-state solutions for $V(\vec{r},t)$ and $T(\vec{r},t)$ are also produced. The electroformed steady-state is reached in approximately 100 ns real time for the present set of initial conditions, boundary conditions, and material properties, consistent with our isothermal phase field method \citep{Sevic2018}.

\begin{figure}
\centering
\includegraphics[scale=0.22, angle=0]{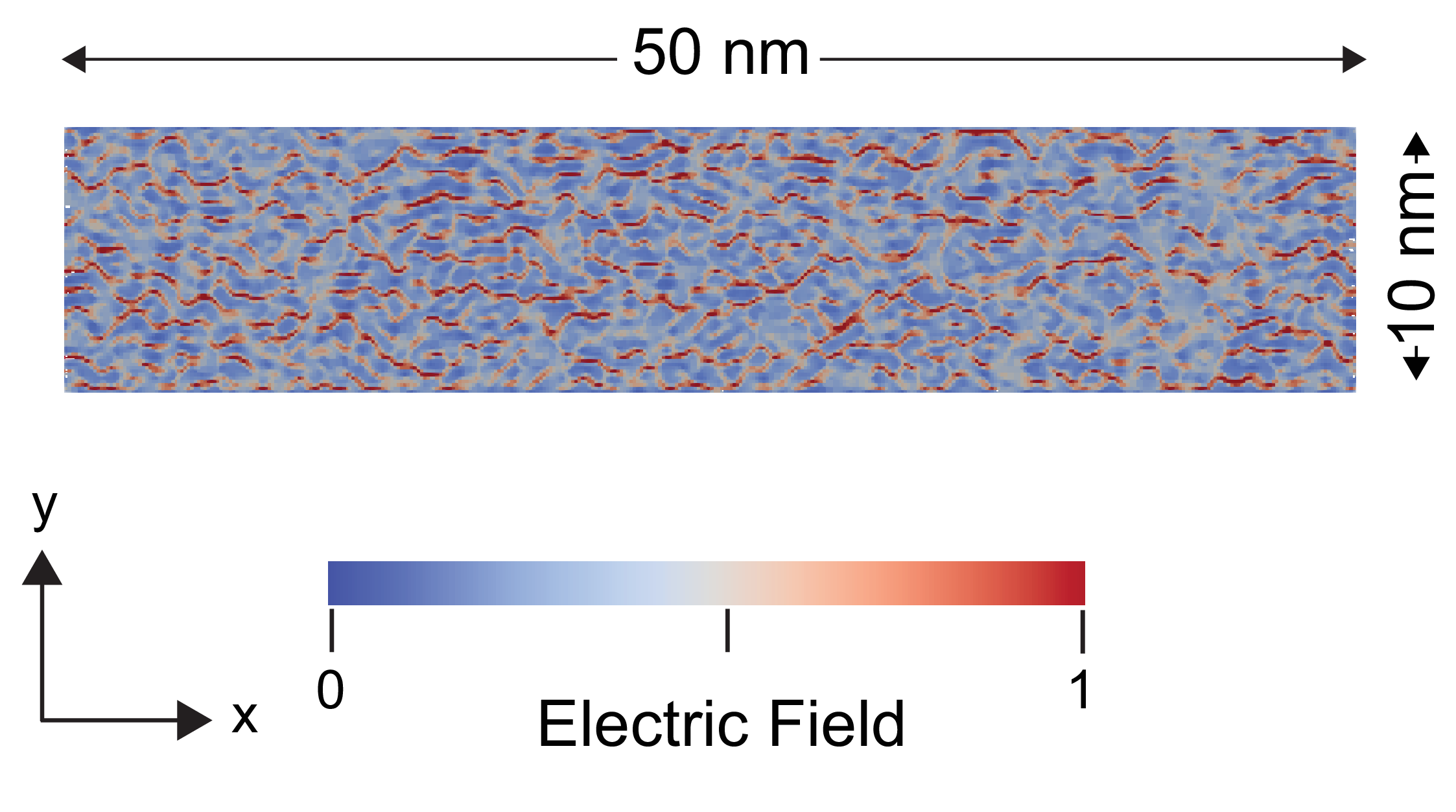}
\caption{Magnitude of normalized electric field immediately following application of the electroforming electric potential at $t = 0$ s, where $|E(\vec{r,t)}| = |\nabla V(\vec{r},t)|$. We assume this initial electric field is established instantaneously, since the charge transport rate our mobility model, Equation \ref{driftActEnergy} is small compared to this transient event. Inhomogeneity in the initial condition of normalized charge density, illustrated by Figure \ref{DeviceStructure}, produces many concentrated clusters of increased electric field intensity, and these regions are uniformly distributed throughout the structure. These field-enhanced thermally excited random clusters are energetically favorable to CF nucleation.}
\label{ElectricFieldInital}
\end{figure}

To understand stochastic CF nucleation and dynamical evolution with our phase field method, consider the vector electric field immediately following the application of the electric potential, illustrated by Figure \ref{ElectricFieldInital}. Here we show electric field magnitude, where $|E(\vec{r,t)}| = |\nabla V(\vec{r},t)|$. We assume this initial electric field is established instantaneously, since the charge transport rate of our mobility model, Equation \ref{driftActEnergy}, is small compared to this transient event \footnote{By immediately following application of the electric potential, we mean one transient time step from $t = 0$ s, approximately 0.1 ns.}\footnote{Showing the magnitude of the vector electric field is reasonable since over the 10 nm x 50 nm scale of our thin film model, $E_{y}(\vec{r},t) >> E_{x}(\vec{r},t)$.}.

It is evident that inhomogeneity in the initial condition of normalized charge density, illustrated by Figure \ref{DeviceStructure}, produces many clusters of increased electric field intensity, and these clusters are uniformly distributed throughout the our thin film. It is these clusters that are energetically favorable to CF nucleation due to the increased local self-heating produced by a concentrated electric field intensity. As local temperature subsequently increases, these various electric field concentration clusters undergo insulator-metal transition to the metallic state, now yielding clusters of increased electrical conductivity. Nevertheless, at this particular instance, these clusters remain distinct and disconnected, and there is no bulk electrical conduction between the top and bottom edges.

\begin{figure}
\centering
\includegraphics[scale=0.22, angle=0]{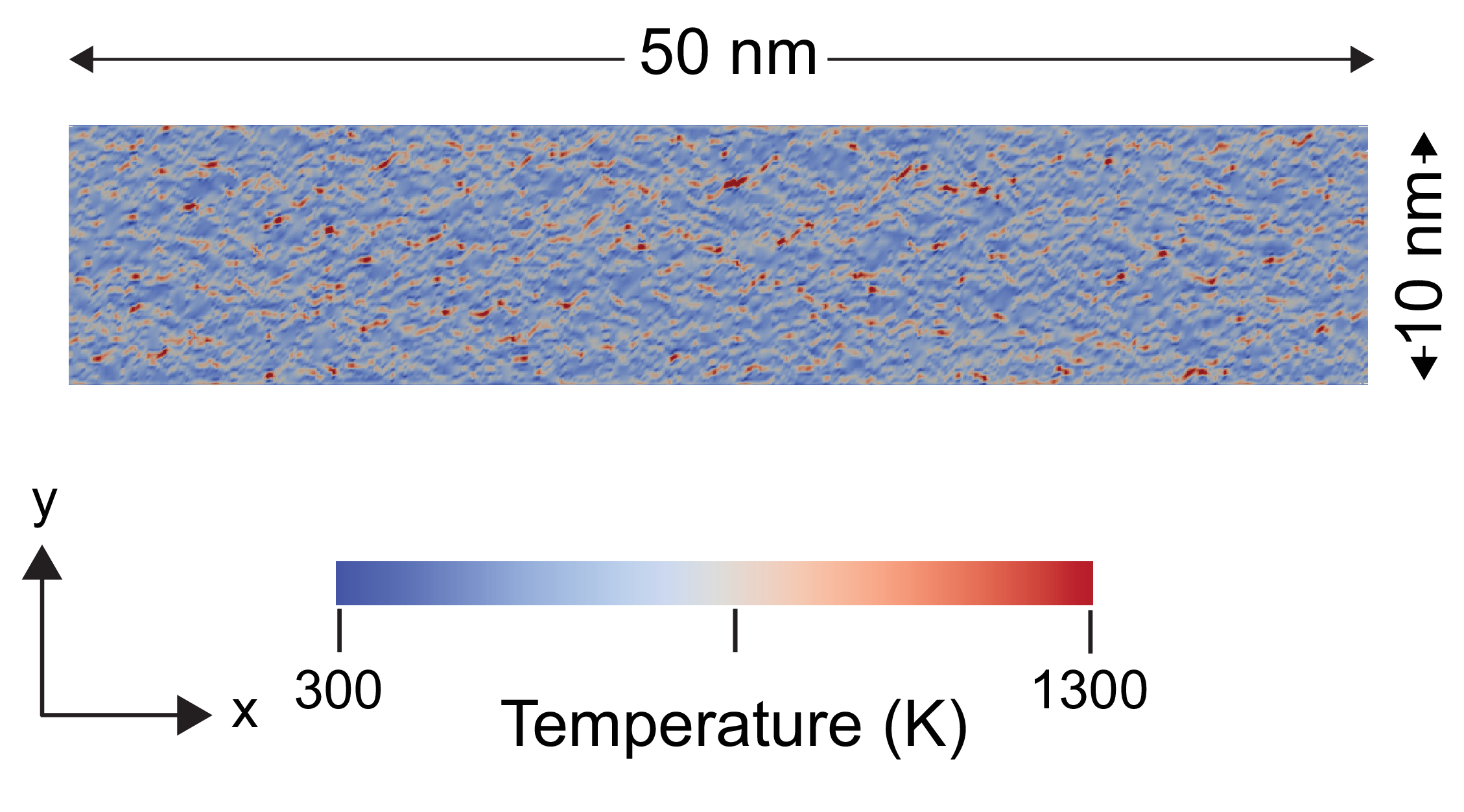}
\caption{Absolute temperature, $T(\vec{r},t)$, shortly after initiation of electroformation, approximately 10 ns in real time for our chosen material properties. It is evident that electrothermal effects are largely inhomogeneous at the outset and that substantial localized Joule heating occurs in correspondence with the initially high local electric field intensity illustrated by Figure \ref{ElectricFieldInital}.}
\label{ThermalFieldInitial}
\end{figure}

Figure \ref{ThermalFieldInitial} shows absolute temperature,  $T(\vec{r},t)$, shortly after initiation of electroformation, approximately 10 ns in real time for our chosen material properties. In concurrence with production of disconnected clusters of concentrated electric field is an associated production of isolated clusters that have reached sufficient absolute temperature for insulator-metal transition to a metallic state, approximately 1300 K for our free-energy density defined by Figure \ref{FEDF}.

Using a variational argument, we may make the following two assertions. First, since these distinct and disconnected clusters now electrically conduct, it is reasonable to conclude that they may longitudinally align to the electric field, $E(\vec{r},t)$, between the top edge and bottom edge of the thin film of Figure \ref{DeviceStructure}, so that disconnected clusters may eventually merge and produce a continuous thread to produce a CF. Second, the foundation of our phase field method, is that part of the total thermodynamic energy of the system is manifested as a diffuse interface between a conducting, metallic state, and an insulating thin film, and energetically favors a minimal interface enclosing these conducting clusters in stationary equilibrium.

\begin{figure}
\centering
\includegraphics[scale=0.22, angle=0]{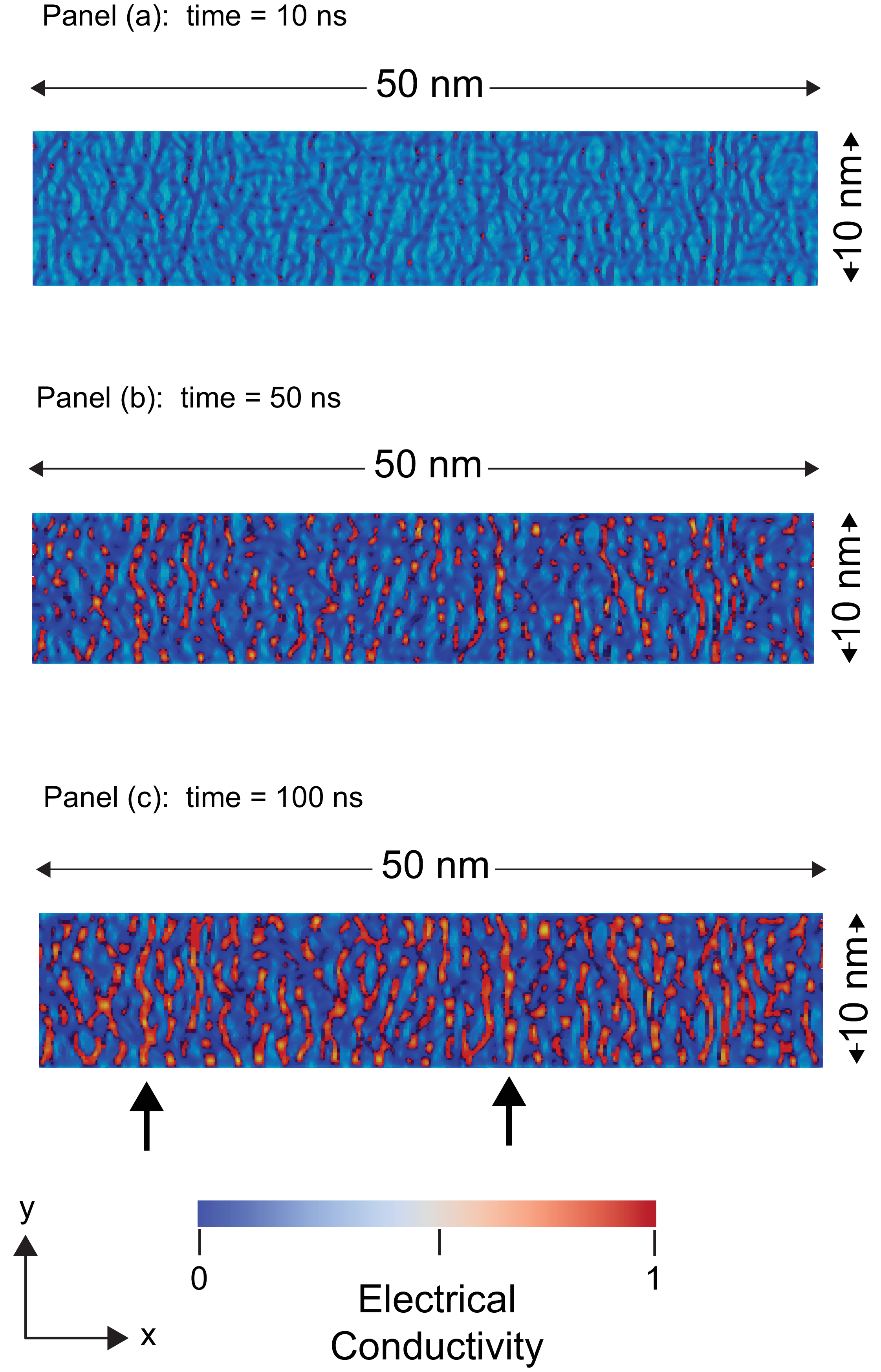}
\caption{Electrical conductivity, $\sigma(\vec{r},c,T)$,at 10 ns, 50 ns and 100 ns in real time. In a sequence of mutually coupled phenomena, CF nucleation is initiated from appropriate initial conditions and material properties, creating isolated clusters in the metallic, conducting, phase. Because it is energetically favorable, some of these isolated clusters eventually merge to create conducting filaments. The two solid arrows on panel (c) point out two continuous conducting filaments traversing the top and bottom edges, each now appropriately electroformed to contribute to bulk resistive switching behavior.}
\label{CurrentDensityTimeSeries}
\end{figure}

In support of this variational argument, consider Figure \ref{CurrentDensityTimeSeries}, illustrating normalized electrical conductivity, $\sigma(\vec{r},c,T)$, at three different times. Panel (a) shows $\sigma(\vec{r},c,T)$ at 10 ns, the same time shown by Figures \ref{ElectricFieldInital} and \ref{ThermalFieldInitial}, illustrating regions of increased electrical conductivity exhibit already characteristic clustering aligned to regions of concentrated Joule heating. Panel (b) shows $\sigma(\vec{r},c,T)$ at 50 ns, illustrating for our model parameters substantial insulator-metal phase transition has occurred, as shown by spontaneously created clusters in the metallic state. Panel (c) shows $\sigma(\vec{r},c,T)$ at 100 ns, the thermodynamic steady-state, representing an electroformed thin film.

Consistent with our variational argument is the spontaenous production of many conducting filaments, embedded within the as-fabricated thin film. In a sequence of mutually coupled phenomena, CF nucleation is initiated from appropriate, stochastic, initial conditions, creating isolated clusters in the metallic, conducting, phase. Because it it energetically favorable to reduce the area of interfaces, some of these isolated conducting clusters eventually merge to create continuous conducting filaments. The conducting filaments produced by our electrothermal phase field method exhibit nanostructure consistent with experimental data \cite{Ahmed2018}\cite{Sun2020}\cite{Zhang2021}.

Not all merged clusters create continuous conducting filaments from the top edge to the bottom edge. Instead, some conducting clusters grow in relative size but nevertheless remain disconnected and unable to contribute to bulk conduction. The two solid arrows shown in Figure \ref{CurrentDensityTimeSeries} illustrate for this particular set of initial conditions and material properties that two complete conducting filaments have formed between the top and bottom edges, each now appropriately electroformed to contribute to bulk resistive switching behavior.

In this letter, we have extended our previous isothermal phase field method studying electroformation by self-consistently coupling to a drift-diffusion transport model of Joule heating \citep{Sevic2019}. Our electrothermal phase field method produced spontaneous nucleation and growth of multiple conducting filaments embedded within an as-fabricated thin film comparable to a range of resistive switches, offering an alternative computational formulation based on metastable atomic-scale states. Our simulation results agree well with data from a range of resistive switches, correctly reproducing experimental conducting filament nanostructure \cite{Ahmed2018}\cite{Sun2020}\cite{Zhang2021}.

\clearpage
\newpage
\bibliography{apssamp}

\providecommand{\noopsort}[1]{}\providecommand{\singleletter}[1]{#1}%
\begin{thebibliography}{35}%
\makeatletter
\providecommand \@ifxundefined [1]{%
 \@ifx{#1\undefined}
}%
\providecommand \@ifnum [1]{%
 \ifnum #1\expandafter \@firstoftwo
 \else \expandafter \@secondoftwo
 \fi
}%
\providecommand \@ifx [1]{%
 \ifx #1\expandafter \@firstoftwo
 \else \expandafter \@secondoftwo
 \fi
}%
\providecommand \natexlab [1]{#1}%
\providecommand \enquote  [1]{``#1''}%
\providecommand \bibnamefont  [1]{#1}%
\providecommand \bibfnamefont [1]{#1}%
\providecommand \citenamefont [1]{#1}%
\providecommand \href@noop [0]{\@secondoftwo}%
\providecommand \href [0]{\begingroup \@sanitize@url \@href}%
\providecommand \@href[1]{\@@startlink{#1}\@@href}%
\providecommand \@@href[1]{\endgroup#1\@@endlink}%
\providecommand \@sanitize@url [0]{\catcode `\\12\catcode `\$12\catcode
  `\&12\catcode `\#12\catcode `\^12\catcode `\_12\catcode `\%12\relax}%
\providecommand \@@startlink[1]{}%
\providecommand \@@endlink[0]{}%
\providecommand \url  [0]{\begingroup\@sanitize@url \@url }%
\providecommand \@url [1]{\endgroup\@href {#1}{\urlprefix }}%
\providecommand \urlprefix  [0]{URL }%
\providecommand \Eprint [0]{\href }%
\providecommand \doibase [0]{https://doi.org/}%
\providecommand \selectlanguage [0]{\@gobble}%
\providecommand \bibinfo  [0]{\@secondoftwo}%
\providecommand \bibfield  [0]{\@secondoftwo}%
\providecommand \translation [1]{[#1]}%
\providecommand \BibitemOpen [0]{}%
\providecommand \bibitemStop [0]{}%
\providecommand \bibitemNoStop [0]{.\EOS\space}%
\providecommand \EOS [0]{\spacefactor3000\relax}%
\providecommand \BibitemShut  [1]{\csname bibitem#1\endcsname}%
\let\auto@bib@innerbib\@empty
\bibitem [{\citenamefont {Strukov}\ \emph {et~al.}(2008)\citenamefont
  {Strukov}, \citenamefont {Snider}, \citenamefont {Stewart},\ and\
  \citenamefont {Williams}}]{Strukov2008}%
  \BibitemOpen
  \bibfield  {author} {\bibinfo {author} {\bibfnamefont {D.~B.}\ \bibnamefont
  {Strukov}}, \bibinfo {author} {\bibfnamefont {G.~S.}\ \bibnamefont {Snider}},
  \bibinfo {author} {\bibfnamefont {D.~R.}\ \bibnamefont {Stewart}},\ and\
  \bibinfo {author} {\bibfnamefont {R.~S.}\ \bibnamefont {Williams}},\
  }\bibfield  {title} {\bibinfo {title} {The missing memristor found},\
  }\href@noop {} {\bibfield  {journal} {\bibinfo  {journal} {Nature}\ }\textbf
  {\bibinfo {volume} {453}},\ \bibinfo {pages} {80 EP } (\bibinfo {year}
  {2008})}\BibitemShut {NoStop}%
\bibitem [{\citenamefont {Waser}\ and\ \citenamefont {Aono}(2007)}]{Waser2007}%
  \BibitemOpen
  \bibfield  {author} {\bibinfo {author} {\bibfnamefont {R.}~\bibnamefont
  {Waser}}\ and\ \bibinfo {author} {\bibfnamefont {M.}~\bibnamefont {Aono}},\
  }\bibfield  {title} {\bibinfo {title} {Nanoionics-based resistive switching
  memories},\ }\href {https://doi.org/10.1038/nmat2023} {\bibfield  {journal}
  {\bibinfo  {journal} {Nature materials}\ }\textbf {\bibinfo {volume} {6}},\
  \bibinfo {pages} {833} (\bibinfo {year} {2007})}\BibitemShut {NoStop}%
\bibitem [{\citenamefont {Xu}\ \emph {et~al.}(2008)\citenamefont {Xu},
  \citenamefont {Liu}, \citenamefont {Sun}, \citenamefont {Liu}, \citenamefont
  {Han}, \citenamefont {Wang}, \citenamefont {Han}, \citenamefont {Kang},\ and\
  \citenamefont {Yu}}]{Xu2008}%
  \BibitemOpen
  \bibfield  {author} {\bibinfo {author} {\bibfnamefont {N.}~\bibnamefont
  {Xu}}, \bibinfo {author} {\bibfnamefont {L.}~\bibnamefont {Liu}}, \bibinfo
  {author} {\bibfnamefont {X.}~\bibnamefont {Sun}}, \bibinfo {author}
  {\bibfnamefont {X.}~\bibnamefont {Liu}}, \bibinfo {author} {\bibfnamefont
  {D.}~\bibnamefont {Han}}, \bibinfo {author} {\bibfnamefont {Y.}~\bibnamefont
  {Wang}}, \bibinfo {author} {\bibfnamefont {R.}~\bibnamefont {Han}}, \bibinfo
  {author} {\bibfnamefont {J.}~\bibnamefont {Kang}},\ and\ \bibinfo {author}
  {\bibfnamefont {B.}~\bibnamefont {Yu}},\ }\bibfield  {title} {\bibinfo
  {title} {Characteristics and mechanism of conduction and set process in
  tin-zno-pt resistance switching random-access memories},\ }\href
  {https://doi.org/10.1063/1.2945278} {\bibfield  {journal} {\bibinfo
  {journal} {Applied Physics Letters}\ }\textbf {\bibinfo {volume} {92}},\
  \bibinfo {pages} {232112} (\bibinfo {year} {2008})}\BibitemShut {NoStop}%
\bibitem [{\citenamefont {Pan}\ \emph {et~al.}(2010)\citenamefont {Pan},
  \citenamefont {Chen}, \citenamefont {shun Wang}, \citenamefont {chao Yang},
  \citenamefont {Yang},\ and\ \citenamefont {Zeng}}]{Pan2010}%
  \BibitemOpen
  \bibfield  {author} {\bibinfo {author} {\bibfnamefont {F.}~\bibnamefont
  {Pan}}, \bibinfo {author} {\bibfnamefont {C.}~\bibnamefont {Chen}}, \bibinfo
  {author} {\bibfnamefont {Z.}~\bibnamefont {shun Wang}}, \bibinfo {author}
  {\bibfnamefont {Y.}~\bibnamefont {chao Yang}}, \bibinfo {author}
  {\bibfnamefont {J.}~\bibnamefont {Yang}},\ and\ \bibinfo {author}
  {\bibfnamefont {F.}~\bibnamefont {Zeng}},\ }\bibfield  {title} {\bibinfo
  {title} {Nonvolatile resistive switching memories-characteristics, mechanisms
  and challenges},\ }\href
  {https://doi.org/https://doi.org/10.1016/S1002-0071(12)60001-X} {\bibfield
  {journal} {\bibinfo  {journal} {Progress in Natural Science: Materials
  International}\ }\textbf {\bibinfo {volume} {20}},\ \bibinfo {pages} {1 }
  (\bibinfo {year} {2010})}\BibitemShut {NoStop}%
\bibitem [{\citenamefont {Ielmini}(2011)}]{Ielmini2011}%
  \BibitemOpen
  \bibfield  {author} {\bibinfo {author} {\bibfnamefont {D.}~\bibnamefont
  {Ielmini}},\ }\bibfield  {title} {\bibinfo {title} {Modeling the universal
  set/reset characteristics of bipolar rram by field- and temperature driven
  filament growth},\ }\href@noop {} {\bibfield  {journal} {\bibinfo  {journal}
  {IEEE Trans. Electron Devices}\ }\textbf {\bibinfo {volume} {58}},\ \bibinfo
  {pages} {4309} (\bibinfo {year} {2011})}\BibitemShut {NoStop}%
\bibitem [{\citenamefont {Larentis}\ \emph {et~al.}(2012)\citenamefont
  {Larentis}, \citenamefont {Nardi}, , \citenamefont {Balatti}, \citenamefont
  {Gilmer},\ and\ \citenamefont {Ielmini}}]{Larentis2012}%
  \BibitemOpen
  \bibfield  {author} {\bibinfo {author} {\bibfnamefont {S.}~\bibnamefont
  {Larentis}}, \bibinfo {author} {\bibfnamefont {.~F.}\ \bibnamefont {Nardi}},
  , \bibinfo {author} {\bibfnamefont {S.}~\bibnamefont {Balatti}}, \bibinfo
  {author} {\bibfnamefont {D.}~\bibnamefont {Gilmer}},\ and\ \bibinfo {author}
  {\bibfnamefont {D.}~\bibnamefont {Ielmini}},\ }\bibfield  {title} {\bibinfo
  {title} {Resistive switching by voltage-driven ion migration in bipolar rram
  - part ii: Modeling},\ }\href@noop {} {\bibfield  {journal} {\bibinfo
  {journal} {IEEE Trans. Electron Devices Lett.}\ }\textbf {\bibinfo {volume}
  {59}},\ \bibinfo {pages} {2468} (\bibinfo {year} {2012})}\BibitemShut
  {NoStop}%
\bibitem [{\citenamefont {Nardi}\ \emph {et~al.}(2012)\citenamefont {Nardi},
  \citenamefont {Larentis}, \citenamefont {Balatti}, \citenamefont {Gilmer},\
  and\ \citenamefont {Ielmini}}]{Nardi2012}%
  \BibitemOpen
  \bibfield  {author} {\bibinfo {author} {\bibfnamefont {F.}~\bibnamefont
  {Nardi}}, \bibinfo {author} {\bibfnamefont {S.}~\bibnamefont {Larentis}},
  \bibinfo {author} {\bibfnamefont {S.}~\bibnamefont {Balatti}}, \bibinfo
  {author} {\bibfnamefont {D.}~\bibnamefont {Gilmer}},\ and\ \bibinfo {author}
  {\bibfnamefont {D.}~\bibnamefont {Ielmini}},\ }\bibfield  {title} {\bibinfo
  {title} {Resistive switching by voltage-driven ion migration in bipolar
  rram—part i: Experimental study},\ }\href@noop {} {\bibfield  {journal}
  {\bibinfo  {journal} {IEEE Trans. Electron Devices Lett.}\ }\textbf {\bibinfo
  {volume} {59}},\ \bibinfo {pages} {2461} (\bibinfo {year}
  {2012})}\BibitemShut {NoStop}%
\bibitem [{\citenamefont {Kim}\ \emph {et~al.}(2015)\citenamefont {Kim},
  \citenamefont {Park},\ and\ \citenamefont {Hwang}}]{Kim2015}%
  \BibitemOpen
  \bibfield  {author} {\bibinfo {author} {\bibfnamefont {K.~M.}\ \bibnamefont
  {Kim}}, \bibinfo {author} {\bibfnamefont {T.~H.}\ \bibnamefont {Park}},\ and\
  \bibinfo {author} {\bibfnamefont {C.~S.}\ \bibnamefont {Hwang}},\ }\bibfield
  {title} {\bibinfo {title} {Dual conical conducting filament model in
  resistance switching tio2 thin films},\ }\href@noop {} {\bibfield  {journal}
  {\bibinfo  {journal} {Scientific Reports}\ }\textbf {\bibinfo {volume} {5}},\
  \bibinfo {pages} {7844 EP } (\bibinfo {year} {2015})}\BibitemShut {NoStop}%
\bibitem [{\citenamefont {Sevic}\ and\ \citenamefont
  {Kobayashi}(2017)}]{Sevic2017}%
  \BibitemOpen
  \bibfield  {author} {\bibinfo {author} {\bibfnamefont {J.~F.}\ \bibnamefont
  {Sevic}}\ and\ \bibinfo {author} {\bibfnamefont {N.~P.}\ \bibnamefont
  {Kobayashi}},\ }\bibfield  {title} {\bibinfo {title} {Multi-physics transient
  simulation of monolithic niobium dioxide-tantalum dioxide memristor selector
  structures},\ }\href@noop {} {\bibfield  {journal} {\bibinfo  {journal}
  {Applied Physics Letters}\ }\textbf {\bibinfo {volume} {111}},\ \bibinfo
  {pages} {153107} (\bibinfo {year} {2017})}\BibitemShut {NoStop}%
\bibitem [{\citenamefont {Sevic}\ and\ \citenamefont
  {Kobayashi}(2018)}]{Sevic2018}%
  \BibitemOpen
  \bibfield  {author} {\bibinfo {author} {\bibfnamefont {J.~F.}\ \bibnamefont
  {Sevic}}\ and\ \bibinfo {author} {\bibfnamefont {N.~P.}\ \bibnamefont
  {Kobayashi}},\ }\bibfield  {title} {\bibinfo {title} {Self-consistent
  continuum-based transient simulation of electroformation of niobium oxide
  tantalum dioxide selector-memristor structures},\ }\href@noop {} {\bibfield
  {journal} {\bibinfo  {journal} {Journal of Applied Physics}\ }\textbf
  {\bibinfo {volume} {124}},\ \bibinfo {pages} {164501} (\bibinfo {year}
  {2018})}\BibitemShut {NoStop}%
\bibitem [{\citenamefont {Yang}\ \emph {et~al.}(2009)\citenamefont {Yang},
  \citenamefont {Miao}, \citenamefont {Pickett}, \citenamefont {Ohlberg},
  \citenamefont {Stewart}, \citenamefont {Lau},\ and\ \citenamefont
  {Williams}}]{Yang2009}%
  \BibitemOpen
  \bibfield  {author} {\bibinfo {author} {\bibfnamefont {J.~J.}\ \bibnamefont
  {Yang}}, \bibinfo {author} {\bibfnamefont {F.}~\bibnamefont {Miao}}, \bibinfo
  {author} {\bibfnamefont {M.~D.}\ \bibnamefont {Pickett}}, \bibinfo {author}
  {\bibfnamefont {D.~A.~A.}\ \bibnamefont {Ohlberg}}, \bibinfo {author}
  {\bibfnamefont {D.~R.}\ \bibnamefont {Stewart}}, \bibinfo {author}
  {\bibfnamefont {C.~N.}\ \bibnamefont {Lau}},\ and\ \bibinfo {author}
  {\bibfnamefont {R.~S.}\ \bibnamefont {Williams}},\ }\bibfield  {title}
  {\bibinfo {title} {The mechanism of electroforming of metal oxide memristive
  switches},\ }\href@noop {} {\bibfield  {journal} {\bibinfo  {journal}
  {Nanotechnology}\ }\textbf {\bibinfo {volume} {20}},\ \bibinfo {pages}
  {215201} (\bibinfo {year} {2009})}\BibitemShut {NoStop}%
\bibitem [{\citenamefont {Strachan}\ \emph {et~al.}(2010)\citenamefont
  {Strachan}, \citenamefont {Pickett}, \citenamefont {Yang}, \citenamefont
  {Aloni}, \citenamefont {David~Kilcoyne}, \citenamefont {Medeiros-Ribeiro},\
  and\ \citenamefont {Stanley~Williams}}]{Strachan2010}%
  \BibitemOpen
  \bibfield  {author} {\bibinfo {author} {\bibfnamefont {J.~P.}\ \bibnamefont
  {Strachan}}, \bibinfo {author} {\bibfnamefont {M.~D.}\ \bibnamefont
  {Pickett}}, \bibinfo {author} {\bibfnamefont {J.~J.}\ \bibnamefont {Yang}},
  \bibinfo {author} {\bibfnamefont {S.}~\bibnamefont {Aloni}}, \bibinfo
  {author} {\bibfnamefont {A.~L.}\ \bibnamefont {David~Kilcoyne}}, \bibinfo
  {author} {\bibfnamefont {G.}~\bibnamefont {Medeiros-Ribeiro}},\ and\ \bibinfo
  {author} {\bibfnamefont {R.}~\bibnamefont {Stanley~Williams}},\ }\bibfield
  {title} {\bibinfo {title} {Direct identification of the conducting channels
  in a functioning memristive device},\ }\href
  {https://doi.org/10.1002/adma.201000186} {\bibfield  {journal} {\bibinfo
  {journal} {Advanced Materials}\ }\textbf {\bibinfo {volume} {22}},\ \bibinfo
  {pages} {3573} (\bibinfo {year} {2010})}\BibitemShut {NoStop}%
\bibitem [{\citenamefont {Miao}\ \emph {et~al.}(2011)\citenamefont {Miao},
  \citenamefont {Strachan}, \citenamefont {Yang}, \citenamefont {Zhang},
  \citenamefont {Goldfarb}, \citenamefont {Torrezan}, \citenamefont {Eschbach},
  \citenamefont {Kelley}, \citenamefont {Medeiros-Ribeiro},\ and\ \citenamefont
  {Williams}}]{Miao2011}%
  \BibitemOpen
  \bibfield  {author} {\bibinfo {author} {\bibfnamefont {F.}~\bibnamefont
  {Miao}}, \bibinfo {author} {\bibfnamefont {J.~P.}\ \bibnamefont {Strachan}},
  \bibinfo {author} {\bibfnamefont {J.~J.}\ \bibnamefont {Yang}}, \bibinfo
  {author} {\bibfnamefont {M.-X.}\ \bibnamefont {Zhang}}, \bibinfo {author}
  {\bibfnamefont {I.}~\bibnamefont {Goldfarb}}, \bibinfo {author}
  {\bibfnamefont {A.~C.}\ \bibnamefont {Torrezan}}, \bibinfo {author}
  {\bibfnamefont {P.}~\bibnamefont {Eschbach}}, \bibinfo {author}
  {\bibfnamefont {R.~D.}\ \bibnamefont {Kelley}}, \bibinfo {author}
  {\bibfnamefont {G.}~\bibnamefont {Medeiros-Ribeiro}},\ and\ \bibinfo {author}
  {\bibfnamefont {R.~S.}\ \bibnamefont {Williams}},\ }\bibfield  {title}
  {\bibinfo {title} {Anatomy of a nanoscale conduction channel reveals the
  mechanism of a high performance memristor},\ }\href
  {https://doi.org/10.1002/adma.201103379} {\bibfield  {journal} {\bibinfo
  {journal} {Advanced Materials}\ }\textbf {\bibinfo {volume} {23}},\ \bibinfo
  {pages} {5633} (\bibinfo {year} {2011})}\BibitemShut {NoStop}%
\bibitem [{\citenamefont {Sun}\ \emph {et~al.}(2014)\citenamefont {Sun},
  \citenamefont {Liu}, \citenamefont {Li}, \citenamefont {Long}, \citenamefont
  {Lv}, \citenamefont {Bi}, \citenamefont {Huo}, \citenamefont {Li},\ and\
  \citenamefont {Liu}}]{Sun2014}%
  \BibitemOpen
  \bibfield  {author} {\bibinfo {author} {\bibfnamefont {H.}~\bibnamefont
  {Sun}}, \bibinfo {author} {\bibfnamefont {Q.}~\bibnamefont {Liu}}, \bibinfo
  {author} {\bibfnamefont {C.}~\bibnamefont {Li}}, \bibinfo {author}
  {\bibfnamefont {S.}~\bibnamefont {Long}}, \bibinfo {author} {\bibfnamefont
  {H.}~\bibnamefont {Lv}}, \bibinfo {author} {\bibfnamefont {C.}~\bibnamefont
  {Bi}}, \bibinfo {author} {\bibfnamefont {Z.}~\bibnamefont {Huo}}, \bibinfo
  {author} {\bibfnamefont {L.}~\bibnamefont {Li}},\ and\ \bibinfo {author}
  {\bibfnamefont {M.}~\bibnamefont {Liu}},\ }\bibfield  {title} {\bibinfo
  {title} {Direct observation of conversion between threshold switching and
  memory switching induced by conductive filament morphology},\ }\href
  {https://doi.org/10.1002/adfm.201401304} {\bibfield  {journal} {\bibinfo
  {journal} {Advanced Functional Materials}\ }\textbf {\bibinfo {volume}
  {24}},\ \bibinfo {pages} {5679} (\bibinfo {year} {2014})}\BibitemShut
  {NoStop}%
\bibitem [{\citenamefont {Ahmed}\ \emph {et~al.}(2018)\citenamefont {Ahmed},
  \citenamefont {Walia}, \citenamefont {Mayes}, \citenamefont {Ramanathan},
  \citenamefont {Guagliardo}, \citenamefont {Bansal}, \citenamefont
  {Bhaskaran}, \citenamefont {Yang},\ and\ \citenamefont {Sriram}}]{Ahmed2018}%
  \BibitemOpen
  \bibfield  {author} {\bibinfo {author} {\bibfnamefont {T.}~\bibnamefont
  {Ahmed}}, \bibinfo {author} {\bibfnamefont {S.}~\bibnamefont {Walia}},
  \bibinfo {author} {\bibfnamefont {E.~L.}\ \bibnamefont {Mayes}}, \bibinfo
  {author} {\bibfnamefont {R.}~\bibnamefont {Ramanathan}}, \bibinfo {author}
  {\bibfnamefont {P.}~\bibnamefont {Guagliardo}}, \bibinfo {author}
  {\bibfnamefont {V.}~\bibnamefont {Bansal}}, \bibinfo {author} {\bibfnamefont
  {M.}~\bibnamefont {Bhaskaran}}, \bibinfo {author} {\bibfnamefont {J.~J.}\
  \bibnamefont {Yang}},\ and\ \bibinfo {author} {\bibfnamefont
  {S.}~\bibnamefont {Sriram}},\ }\bibfield  {title} {\bibinfo {title} {Inducing
  tunable switching behavior in a single memristor},\ }\href
  {https://doi.org/https://doi.org/10.1016/j.apmt.2018.03.003} {\bibfield
  {journal} {\bibinfo  {journal} {Applied Materials Today}\ }\textbf {\bibinfo
  {volume} {11}},\ \bibinfo {pages} {280 } (\bibinfo {year}
  {2018})}\BibitemShut {NoStop}%
\bibitem [{\citenamefont {Sun}\ \emph {et~al.}(2020)\citenamefont {Sun},
  \citenamefont {Han}, \citenamefont {Xu},\ and\ \citenamefont
  {Qian}}]{Sun2020}%
  \BibitemOpen
  \bibfield  {author} {\bibinfo {author} {\bibfnamefont {B.}~\bibnamefont
  {Sun}}, \bibinfo {author} {\bibfnamefont {X.}~\bibnamefont {Han}}, \bibinfo
  {author} {\bibfnamefont {R.}~\bibnamefont {Xu}},\ and\ \bibinfo {author}
  {\bibfnamefont {K.}~\bibnamefont {Qian}},\ }\bibfield  {title} {\bibinfo
  {title} {Uncovering the indium filament formation and dissolution in
  transparent ito/sinx/ito resistive random access memory},\ }\href
  {https://doi.org/10.1021/acsaelm.0c00193} {\bibfield  {journal} {\bibinfo
  {journal} {ACS Applied Electronic Materials}\ }\textbf {\bibinfo {volume}
  {2}},\ \bibinfo {pages} {1603} (\bibinfo {year} {2020})}\BibitemShut
  {NoStop}%
\bibitem [{\citenamefont {Zhang}(2021)}]{Zhang2021}%
  \BibitemOpen
  \bibfield  {author} {\bibinfo {author} {\bibfnamefont {Y.}~\bibnamefont
  {Zhang}},\ }\bibfield  {title} {\bibinfo {title} {Evolution of the conductive
  filament system in hfo2 based memristors observed by direct atomic-scale
  imaging},\ }\href {https://doi.org/doi.org/10.1038/s41467-021-27575-z}
  {\bibfield  {journal} {\bibinfo  {journal} {Nat Commun}\ }\textbf {\bibinfo
  {volume} {12}},\ \bibinfo {pages} {7232} (\bibinfo {year}
  {2021})}\BibitemShut {NoStop}%
\bibitem [{\citenamefont {Sevic}\ and\ \citenamefont
  {Kobayashi}(2019)}]{Sevic2019}%
  \BibitemOpen
  \bibfield  {author} {\bibinfo {author} {\bibfnamefont {J.~F.}\ \bibnamefont
  {Sevic}}\ and\ \bibinfo {author} {\bibfnamefont {N.~P.}\ \bibnamefont
  {Kobayashi}},\ }\bibfield  {title} {\bibinfo {title} {A computational phase
  field study of conducting channel formation in dielectric thin films: A view
  toward the physical origins of resistive switching},\ }\href
  {https://doi.org/10.1063/1.5110911} {\bibfield  {journal} {\bibinfo
  {journal} {Journal of Applied Physics}\ }\textbf {\bibinfo {volume} {126}},\
  \bibinfo {pages} {065305} (\bibinfo {year} {2019})}\BibitemShut {NoStop}%
\bibitem [{Note1()}]{Note1}%
  \BibitemOpen
  \bibinfo {note} {By pristine we mean pre-electroformed}\BibitemShut {NoStop}%
\bibitem [{Note2()}]{Note2}%
  \BibitemOpen
  \bibinfo {note} {Spatial and time dependence of the three state variables,
  $c(\protect \vec {r},t)$, $T(\protect \vec {r},t)$, and $V(\protect \vec
  {r},t)$, is always implicit, and occasionally may be suppressed subsequently
  due to space limitations.}\BibitemShut {Stop}%
\bibitem [{\citenamefont {Leon}\ \emph {et~al.}(2017)\citenamefont {Leon},
  \citenamefont {Norris}, \citenamefont {Yang}, \citenamefont {Sevic},\ and\
  \citenamefont {Kobayashi}}]{DiazLeon2017}%
  \BibitemOpen
  \bibfield  {author} {\bibinfo {author} {\bibfnamefont {J.~J.~D.}\
  \bibnamefont {Leon}}, \bibinfo {author} {\bibfnamefont {K.~J.}\ \bibnamefont
  {Norris}}, \bibinfo {author} {\bibfnamefont {J.~J.}\ \bibnamefont {Yang}},
  \bibinfo {author} {\bibfnamefont {J.~F.}\ \bibnamefont {Sevic}},\ and\
  \bibinfo {author} {\bibfnamefont {N.~P.}\ \bibnamefont {Kobayashi}},\
  }\bibfield  {title} {\bibinfo {title} {A niobium oxide-tantalum oxide
  selector memristor self-aligned nanostack},\ }\href@noop {} {\bibfield
  {journal} {\bibinfo  {journal} {Applied Physics Letters}\ }\textbf {\bibinfo
  {volume} {110}},\ \bibinfo {pages} {103102} (\bibinfo {year}
  {2017})}\BibitemShut {NoStop}%
\bibitem [{\citenamefont {Leon}\ \emph {et~al.}(2016)\citenamefont {Leon},
  \citenamefont {Norris}, \citenamefont {Yang}, \citenamefont {Sevic},\ and\
  \citenamefont {Kobayashi}}]{DiazLeon2016}%
  \BibitemOpen
  \bibfield  {author} {\bibinfo {author} {\bibfnamefont {J.~J.~D.}\
  \bibnamefont {Leon}}, \bibinfo {author} {\bibfnamefont {K.~J.}\ \bibnamefont
  {Norris}}, \bibinfo {author} {\bibfnamefont {J.~J.}\ \bibnamefont {Yang}},
  \bibinfo {author} {\bibfnamefont {J.~F.}\ \bibnamefont {Sevic}},\ and\
  \bibinfo {author} {\bibfnamefont {N.~P.}\ \bibnamefont {Kobayashi}},\
  }\bibfield  {title} {\bibinfo {title} {Integration of a niobium oxide
  selector on a tantalum oxide memristor by local oxidation using joule
  heating},\ }\href@noop {} {\bibfield  {journal} {\bibinfo  {journal} {SPIE,
  San Diego}\ } (\bibinfo {year} {2016})}\BibitemShut {NoStop}%
\bibitem [{Note3()}]{Note3}%
  \BibitemOpen
  \bibinfo {note} {For the current formulation, $\Omega $ scales the volume of
  a mesh cell, of unit thcikness.}\BibitemShut {Stop}%
\bibitem [{\citenamefont {Querre}\ \emph {et~al.}(2018)\citenamefont {Querre},
  \citenamefont {Tranchant}, \citenamefont {Corraze}, \citenamefont {Cordier},
  \citenamefont {Bouquet}, \citenamefont {Députier}, \citenamefont
  {Guilloux-Viry}, \citenamefont {Besland}, \citenamefont {Janod},\ and\
  \citenamefont {Cario}}]{Querre2018}%
  \BibitemOpen
  \bibfield  {author} {\bibinfo {author} {\bibfnamefont {M.}~\bibnamefont
  {Querre}}, \bibinfo {author} {\bibfnamefont {J.}~\bibnamefont {Tranchant}},
  \bibinfo {author} {\bibfnamefont {B.}~\bibnamefont {Corraze}}, \bibinfo
  {author} {\bibfnamefont {S.}~\bibnamefont {Cordier}}, \bibinfo {author}
  {\bibfnamefont {V.}~\bibnamefont {Bouquet}}, \bibinfo {author} {\bibfnamefont
  {S.}~\bibnamefont {Députier}}, \bibinfo {author} {\bibfnamefont
  {M.}~\bibnamefont {Guilloux-Viry}}, \bibinfo {author} {\bibfnamefont
  {M.}~\bibnamefont {Besland}}, \bibinfo {author} {\bibfnamefont
  {E.}~\bibnamefont {Janod}},\ and\ \bibinfo {author} {\bibfnamefont
  {L.}~\bibnamefont {Cario}},\ }\bibfield  {title} {\bibinfo {title}
  {Non-volatile resistive switching in the mott insulator (v1-xcrx)(2)o-3},\
  }\href@noop {} {\bibfield  {journal} {\bibinfo  {journal} {Physica B:
  Condensed Matter}\ } (\bibinfo {year} {2018})}\BibitemShut {NoStop}%
\bibitem [{\citenamefont {Tranchant}\ \emph {et~al.}(2018)\citenamefont
  {Tranchant}, \citenamefont {Querre}, \citenamefont {Janod}, \citenamefont
  {Besland}, \citenamefont {Corraze},\ and\ \citenamefont
  {Cario}}]{Tranchant2018}%
  \BibitemOpen
  \bibfield  {author} {\bibinfo {author} {\bibfnamefont {J.}~\bibnamefont
  {Tranchant}}, \bibinfo {author} {\bibfnamefont {M.}~\bibnamefont {Querre}},
  \bibinfo {author} {\bibfnamefont {E.}~\bibnamefont {Janod}}, \bibinfo
  {author} {\bibfnamefont {M.}~\bibnamefont {Besland}}, \bibinfo {author}
  {\bibfnamefont {B.}~\bibnamefont {Corraze}},\ and\ \bibinfo {author}
  {\bibfnamefont {L.}~\bibnamefont {Cario}},\ }\bibfield  {title} {\bibinfo
  {title} {Mott memory devices based on the mott insulator (v1-xcrx)2o3},\
  }\href@noop {} {\bibfield  {journal} {\bibinfo  {journal} {IEEE 2018
  International Memory Workshop}\ } (\bibinfo {year} {2018})}\BibitemShut
  {NoStop}%
\bibitem [{\citenamefont {Provatas}\ and\ \citenamefont
  {Elder}(2010)}]{Provatas2010}%
  \BibitemOpen
  \bibfield  {author} {\bibinfo {author} {\bibfnamefont {N.}~\bibnamefont
  {Provatas}}\ and\ \bibinfo {author} {\bibfnamefont {K.}~\bibnamefont
  {Elder}},\ }\href@noop {} {\emph {\bibinfo {title} {Phase-Field Methods in
  Materials Science and Engineering}}}\ (\bibinfo  {publisher} {Wiley-VCH
  Verlag},\ \bibinfo {address} {Weinheim, Germany},\ \bibinfo {year}
  {2010})\BibitemShut {NoStop}%
\bibitem [{\citenamefont {Cahn}\ and\ \citenamefont
  {Hilliard}(1958)}]{Cahn1958}%
  \BibitemOpen
  \bibfield  {author} {\bibinfo {author} {\bibfnamefont {J.~W.}\ \bibnamefont
  {Cahn}}\ and\ \bibinfo {author} {\bibfnamefont {J.~E.}\ \bibnamefont
  {Hilliard}},\ }\bibfield  {title} {\bibinfo {title} {Free energy of a
  nonuniform system. i. interfacial free energy},\ }\href
  {https://doi.org/10.1063/1.1744102} {\bibfield  {journal} {\bibinfo
  {journal} {The Journal of Chemical Physics}\ }\textbf {\bibinfo {volume}
  {28}},\ \bibinfo {pages} {258} (\bibinfo {year} {1958})}\BibitemShut
  {NoStop}%
\bibitem [{\citenamefont {Kim}\ \emph {et~al.}(1998)\citenamefont {Kim},
  \citenamefont {Kim},\ and\ \citenamefont {Suzuki}}]{KKS}%
  \BibitemOpen
  \bibfield  {author} {\bibinfo {author} {\bibfnamefont {S.~G.}\ \bibnamefont
  {Kim}}, \bibinfo {author} {\bibfnamefont {W.~T.}\ \bibnamefont {Kim}},\ and\
  \bibinfo {author} {\bibfnamefont {T.}~\bibnamefont {Suzuki}},\ }\bibfield
  {title} {\bibinfo {title} {Interfacial compositions of solid and liquid in a
  phase field model with finite interface thickness for isothermal
  solidification in binary alloys},\ }\href@noop {} {\bibfield  {journal}
  {\bibinfo  {journal} {Phys. Rev. E}\ }\textbf {\bibinfo {volume} {58}},\
  \bibinfo {pages} {3316} (\bibinfo {year} {1998})}\BibitemShut {NoStop}%
\bibitem [{\citenamefont {Gaston}\ \emph {et~al.}(2009)\citenamefont {Gaston},
  \citenamefont {Newman}, \citenamefont {Hansen},\ and\ \citenamefont
  {Lebrun-Grandi{\'e}}}]{gaston2009moose}%
  \BibitemOpen
  \bibfield  {author} {\bibinfo {author} {\bibfnamefont {D.}~\bibnamefont
  {Gaston}}, \bibinfo {author} {\bibfnamefont {C.}~\bibnamefont {Newman}},
  \bibinfo {author} {\bibfnamefont {G.}~\bibnamefont {Hansen}},\ and\ \bibinfo
  {author} {\bibfnamefont {D.}~\bibnamefont {Lebrun-Grandi{\'e}}},\ }\bibfield
  {title} {\bibinfo {title} {Moose: {A} parallel computational framework for
  coupled systems of nonlinear equations},\ }\href@noop {} {\bibfield
  {journal} {\bibinfo  {journal} {Nuclear Engineering and Design}\ }\textbf
  {\bibinfo {volume} {239}},\ \bibinfo {pages} {1768} (\bibinfo {year}
  {2009})}\BibitemShut {NoStop}%
\bibitem [{\citenamefont {Tonks}\ \emph {et~al.}(2012)\citenamefont {Tonks},
  \citenamefont {Gaston}, \citenamefont {Millett}, \citenamefont {Andrs},\ and\
  \citenamefont {Talbot}}]{Tonks2010}%
  \BibitemOpen
  \bibfield  {author} {\bibinfo {author} {\bibfnamefont {M.~R.}\ \bibnamefont
  {Tonks}}, \bibinfo {author} {\bibfnamefont {D.}~\bibnamefont {Gaston}},
  \bibinfo {author} {\bibfnamefont {P.~C.}\ \bibnamefont {Millett}}, \bibinfo
  {author} {\bibfnamefont {D.}~\bibnamefont {Andrs}},\ and\ \bibinfo {author}
  {\bibfnamefont {P.}~\bibnamefont {Talbot}},\ }\bibfield  {title} {\bibinfo
  {title} {An object-oriented finite element framework for multiphysics phase
  field simulations},\ }\href@noop {} {\bibfield  {journal} {\bibinfo
  {journal} {Computational Materials Science}\ }\textbf {\bibinfo {volume}
  {1}},\ \bibinfo {pages} {20 } (\bibinfo {year} {2012})}\BibitemShut {NoStop}%
\bibitem [{\citenamefont {Schwen}\ \emph {et~al.}(2017)\citenamefont {Schwen},
  \citenamefont {Aagesen}, \citenamefont {Peterson},\ and\ \citenamefont
  {Tonks}}]{schwen2017}%
  \BibitemOpen
  \bibfield  {author} {\bibinfo {author} {\bibfnamefont {D.}~\bibnamefont
  {Schwen}}, \bibinfo {author} {\bibfnamefont {L.}~\bibnamefont {Aagesen}},
  \bibinfo {author} {\bibfnamefont {J.}~\bibnamefont {Peterson}},\ and\
  \bibinfo {author} {\bibfnamefont {M.}~\bibnamefont {Tonks}},\ }\bibfield
  {title} {\bibinfo {title} {Rapid multiphase-field model development using a
  modular free energy based approach with automatic differentiation in
  moose/marmot},\ }\href@noop {} {\bibfield  {journal} {\bibinfo  {journal}
  {Computational Materials Science}\ }\textbf {\bibinfo {volume} {132}},\
  \bibinfo {pages} {36} (\bibinfo {year} {2017})}\BibitemShut {NoStop}%
\bibitem [{\citenamefont {Balay}\ \emph {et~al.}(2016)\citenamefont {Balay},
  \citenamefont {andMark F.~Adams}, \citenamefont {Brown}, \citenamefont
  {Brune}, \citenamefont {Buschelman}, \citenamefont {Dalcin}, \citenamefont
  {Eijkhout}, \citenamefont {Gropp}, \citenamefont {Kaushik}, \citenamefont
  {Knepley}, \citenamefont {McInnes}, \citenamefont {Rupp}, \citenamefont
  {Smith}, \citenamefont {Zampini}, \citenamefont {Zhang},\ and\ \citenamefont
  {Zhang}}]{petsc}%
  \BibitemOpen
  \bibfield  {author} {\bibinfo {author} {\bibfnamefont {S.}~\bibnamefont
  {Balay}}, \bibinfo {author} {\bibfnamefont {S.~A.}\ \bibnamefont {andMark
  F.~Adams}}, \bibinfo {author} {\bibfnamefont {J.}~\bibnamefont {Brown}},
  \bibinfo {author} {\bibfnamefont {P.}~\bibnamefont {Brune}}, \bibinfo
  {author} {\bibfnamefont {K.}~\bibnamefont {Buschelman}}, \bibinfo {author}
  {\bibfnamefont {L.}~\bibnamefont {Dalcin}}, \bibinfo {author} {\bibfnamefont
  {V.}~\bibnamefont {Eijkhout}}, \bibinfo {author} {\bibfnamefont {W.~D.}\
  \bibnamefont {Gropp}}, \bibinfo {author} {\bibfnamefont {D.}~\bibnamefont
  {Kaushik}}, \bibinfo {author} {\bibfnamefont {M.~G.}\ \bibnamefont
  {Knepley}}, \bibinfo {author} {\bibfnamefont {L.~C.}\ \bibnamefont
  {McInnes}}, \bibinfo {author} {\bibfnamefont {K.}~\bibnamefont {Rupp}},
  \bibinfo {author} {\bibfnamefont {B.~F.}\ \bibnamefont {Smith}}, \bibinfo
  {author} {\bibfnamefont {S.}~\bibnamefont {Zampini}}, \bibinfo {author}
  {\bibfnamefont {H.}~\bibnamefont {Zhang}},\ and\ \bibinfo {author}
  {\bibfnamefont {H.}~\bibnamefont {Zhang}},\ }\href@noop {} {\emph {\bibinfo
  {title} {{PETS}c Users Manual}}},\ \bibinfo {type} {Tech. Rep.}\ \bibinfo
  {number} {ANL-95/11 - Revision 3.7}\ (\bibinfo  {institution} {Argonne
  National Laboratory},\ \bibinfo {year} {2016})\BibitemShut {NoStop}%
\bibitem [{\citenamefont {Kirk}\ \emph {et~al.}(2006)\citenamefont {Kirk},
  \citenamefont {Peterson}, \citenamefont {Stogner},\ and\ \citenamefont
  {Carey}}]{libMeshPaper}%
  \BibitemOpen
  \bibfield  {author} {\bibinfo {author} {\bibfnamefont {B.~S.}\ \bibnamefont
  {Kirk}}, \bibinfo {author} {\bibfnamefont {J.~W.}\ \bibnamefont {Peterson}},
  \bibinfo {author} {\bibfnamefont {R.~H.}\ \bibnamefont {Stogner}},\ and\
  \bibinfo {author} {\bibfnamefont {G.~F.}\ \bibnamefont {Carey}},\ }\bibfield
  {title} {\bibinfo {title} {{\texttt{libMesh}: A C++ Library for Parallel
  Adaptive Mesh Refinement/Coarsening Simulations}},\ }\href@noop {} {\bibfield
   {journal} {\bibinfo  {journal} {Engineering with Computers}\ }\textbf
  {\bibinfo {volume} {22}},\ \bibinfo {pages} {237} (\bibinfo {year}
  {2006})}\BibitemShut {NoStop}%
\bibitem [{Note4()}]{Note4}%
  \BibitemOpen
  \bibinfo {note} {By immediately following application of the electric
  potential, we mean one transient time step from $t = 0$ s, approximately 0.1
  ns.}\BibitemShut {Stop}%
\bibitem [{Note5()}]{Note5}%
  \BibitemOpen
  \bibinfo {note} {Showing the magnitude of the vector electric field is
  reasonable since over the 10 nm x 50 nm scale of our thin film model,
  $E_{y}(\protect \vec {r},t) >> E_{x}(\protect \vec {r},t)$.}\BibitemShut
  {Stop}%
\end{thebibliography}%

\end{document}